\documentclass[aps,pra,twocolumn,superscriptaddress]{revtex4}
\usepackage{amsfonts}
\usepackage{amssymb}
\usepackage{amsmath}
\usepackage{epsfig}
\usepackage{color}
\usepackage{graphics, graphicx}
\usepackage{bbold}
\usepackage{psfrag}
\usepackage{mathcomp}
\usepackage{subfigure}
\usepackage{verbatim}
\usepackage{color}
\usepackage[colorlinks,citecolor=blue]{hyperref}

\begin{document}

\title{Interaction-induced exotic vortex states in an optical lattice clock
with spin-orbit coupling}
\author{Xiaofan Zhou}
\affiliation{State Key Laboratory of Quantum Optics and Quantum Optics Devices, Institute
of Laser spectroscopy, Shanxi University, Taiyuan 030006, China}
\affiliation{Collaborative Innovation Center of Extreme Optics, Shanxi University,
Taiyuan, Shanxi 030006, China}
\author{Jian-Song Pan}
\affiliation{Key Laboratory of Quantum Information, University of Science and Technology
of China, CAS, Hefei, Anhui, 230026, China}
\affiliation{Synergetic Innovation Center of Quantum Information and Quantum Physics,
University of Science and Technology of China, Hefei, Anhui 230026, China}
\affiliation{Wilczek Quantum Center, School of Physics and Astronomy and T. D. Lee
Institute, Shanghai Jiao Tong University, Shanghai 200240, China}
\author{Wei Yi}
\email{wyiz@ustc.edu.cn}
\affiliation{Key Laboratory of Quantum Information, University of Science and Technology
of China, CAS, Hefei, Anhui, 230026, China}
\affiliation{Synergetic Innovation Center of Quantum Information and Quantum Physics,
University of Science and Technology of China, Hefei, Anhui 230026, China}
\author{Gang Chen}
\email{chengang971@163.com}
\affiliation{State Key Laboratory of Quantum Optics and Quantum Optics Devices, Institute
of Laser spectroscopy, Shanxi University, Taiyuan 030006, China}
\affiliation{Collaborative Innovation Center of Extreme Optics, Shanxi University,
Taiyuan, Shanxi 030006, China}
\author{Suotang Jia}
\affiliation{State Key Laboratory of Quantum Optics and Quantum Optics Devices, Institute
of Laser spectroscopy, Shanxi University, Taiyuan 030006, China}
\affiliation{Collaborative Innovation Center of Extreme Optics, Shanxi University,
Taiyuan, Shanxi 030006, China}

\begin{abstract}
Motivated by a recent experiment [L. F. Livi, \textit{et al.}, Phys. Rev.
Lett.~\textbf{117}, 220401(2016)], we study the ground-state properties of
interacting fermions in a one-dimensional optical lattice clock with
spin-orbit coupling. As the electronic and the hyperfine-spin states in the
clock-state manifolds can be treated as effective sites along distinct
synthetic dimensions, the system can be considered as multiple two-leg
ladders with uniform magnetic flux penetrating the plaquettes of each
ladder. As the inter-orbital spin-exchange interactions in the clock-state
manifolds couple individual ladders together, we show that exotic
interaction-induced vortex states emerge in the coupled-ladder system, which
compete with existing phases of decoupled ladders and lead to a rich phase
diagram. Adopting the density matrix renormalization group approach, we map
out the phase diagram, and investigate in detail the currents and the
density-density correlations of the various phases. Our results reveal the
impact of interactions on spin-orbit coupled systems, and are particularly
relevant to the on-going exploration of spin-orbit coupled optical lattice
clocks.
\end{abstract}

\maketitle


\section{Introduction}

Spin-orbit coupling (SOC) plays a key role in solid-state topological
materials such as topological insulators and quantum spin Hall systems~\cite%
{kanereview,zhangscreview,alicea}. The experimental realization of synthetic
SOC in cold atomic gases opens up the avenue of simulating synthetic
topological matter on the versatile platform of cold atoms~\cite%
{SOC1,SOC2,SOC3,2dsoc1,2dsoc2,2dsoc3,clock173Yb,clock87Sr,socreview1,socreview2,socreview3,socreview4,socreview5,socreview6,socreview7}%
. In most of the previous studies, synthetic SOC is typically implemented in
alkali atoms using a two-photon Raman process, in which different spin
states in the ground-state hyperfine manifold of the atoms are coupled.
Thus, as the atoms undergo Raman-assisted spin flips, their center-of-mass
momenta also change due to the photon recoil. Alternatively, by considering
the atomic spin states as discrete lattice sites, the SOC can also be mapped
to effective tunneling in the so-called synthetic dimension~\cite%
{SDprl12,SDprl14,SDnjp15,SDpra16}. Such an interpretation has led to the
realization of two-leg ladder models with synthetic magnetic flux, and to
the subsequent experimental demonstration of chiral edge states using cold
atoms~\cite{clock173Yb,RamanRb,AEnew8}.

For systems under the synthetic SOC generated by the Raman scheme, a key
experimental difficulty in reaching the desired many-body ground states is
the heating caused by high-lying excited states in the Raman process, whose
single-photon detuning is limited by the fine-structure splitting~\cite%
{SOC1,SOC2,SOC3}. This problem can be overcome either by choosing atomic
species with large fine-structure splitting~\cite{lanth1,lanth2}, or by
using alkaline-earth-like atoms~\cite{clocktheory}, which feature long-lived
excited states. Indeed, in two recent experiments, synthetic SOCs with
significantly reduced heating have been experimentally demonstrated by
directly coupling the ground $^1S_0$ (referred to as $|g\rangle$) and the
metastable $^3P_0$ (referred to as $|e\rangle$) clock-state manifolds of $%
^{87}$Sr or $^{173}$Yb lattice clocks~\cite{clock173Yb,clock87Sr}. In Ref.~%
\cite{clock173Yb}, the electronic states are further mapped onto the
effective lattice sites along a synthetic dimension, such that a two-leg
ladder model with uniform magnetic flux is realized and the resulting chiral
edge currents are probed. Similar models for bosonic systems have been
extensively investigated in the past~\cite%
{Boseladder0,Boseladder1,Boseladder2,Boseladder3,Boseladder4,Boseladder5,Boseladder6,Boseladder7}%
. When the flux is small, chiral edge currents emerge at the system
boundary, where the currents along the two legs are opposite in direction.
This is reminiscent of the Meissner effects of superconductivity~\cite%
{Meissnersuperconductors}. When the flux becomes sufficiently large, the
system undergoes a phase transition as the chiral edge currents are replaced
by vortex lattices in the bulk, where currents exist on both the rungs and
the edges of the ladder.

Furthermore, when taking the hyperfine spin states in the clock-state
manifolds into account, one can map both the electronic and the spin degrees
of freedom into distinct synthetic dimensions, such that the system in Ref.~%
\cite{clock173Yb} can be extended to model multiple two-leg ladders with
synthetic magnetic flux penetrating the plaquettes of each ladder. Here,
different electronic states label the two legs of each ladder, and different
spin states label different ladders. It would then be interesting to study
the ground state of the system as the parameters such as the flux or the
interactions are tuned. In the clock-state manifolds of alkaline-earth-like
atoms, the nuclear and the electronic degrees of freedom are separated, and
the short-range two-body interactions occur either in the electronic
spin-singlet channel $|-\rangle =\frac{1}{2}(|ge\rangle -|eg\rangle )\otimes
(|\!\!\downarrow \uparrow \rangle +|\!\!\uparrow \downarrow \rangle )$, or
in the electronic spin-triplet channel $|+\rangle =\frac{1}{2}(|ge\rangle
+|eg\rangle )\otimes (|\!\!\downarrow \uparrow \rangle -|\!\!\uparrow
\downarrow \rangle )$~\cite{ofr1,ofr2,ofr3}. Here, $|\!\!\!\uparrow \rangle $
and $|\!\!\!\downarrow \rangle $ label different spin states in the $%
|g\rangle $ or $|e\rangle $ hyperfine manifolds. As reported in previous
studies, such an inter-orbital spin-exchange interaction would induce
density-ordered states in either the spin or the charge channel, leading to
spin-density wave (SDW), orbital-density order (ODW) or charge-density wave
(CDW) phases~\cite%
{phasePRL,phaseEPL,quella13,twoorbitSUn1,quella15,twoorbitSUn2,twoorbitSUnreview,zhou,Zoller}%
. More interestingly, these interactions would couple the otherwise
independent ladders, which may induce new patterns of current flow in the
system.

In this work, adopting the concept of synthetic dimensions, we explicitly
consider the hyperfine spin states in the clock-state manifolds and map the
system in Ref.~\cite{clock173Yb} to multiple two-leg ladders (see Fig.~%
\ref{experimental}). Using the density matrix renormalization group (DMRG)
approach~\cite{dmrg1,dmrg2}, we then numerically investigate the effect of
interactions on the many-body ground-state properties such as the current
flow and the density-density correlations. Our numerical results reveal a
rich phase diagram, where the interactions drastically modify the Meissner
and the vortex states in the non-interacting case. In particular, we show
the existence of an exotic interaction-induced vortex state, where spin
currents emerge between different ladders together with SDW in the system.
As the interactions in the clock-state manifolds can be readily tuned by
external magnetic field through the orbital Feshbach resonance~\cite%
{ren1,ofrexp1,ofrexp2}, or by transverse trapping frequencies through the
confinement-induced resonance~\cite{CIR,ren2}, our results have interesting
implications for future experiments.

The work is organized as follows. In Sec.~\ref{Model and Hamiltonian}, we
present the system setup and the mode Hamiltonian. We discuss the phase
diagram of the non-interacting case in Sec.~\ref{Phases and currents in the
non-interacting case}. We then study in detail the impact of interactions on
the Meissner state and the vortex state, respectively in Secs.~\ref{Impact
of interactions on the Meissner state} and~\ref{Impact of interactions on
the vortex state}. The detection and conclusion are given respectively in Secs.~\ref%
{Discussions} and~\ref{Conclusion}.\newline

\section{Model and Hamiltonian}

\label{Model and Hamiltonian}

\begin{figure}[tb]
\centering
\includegraphics[width =8.5cm]{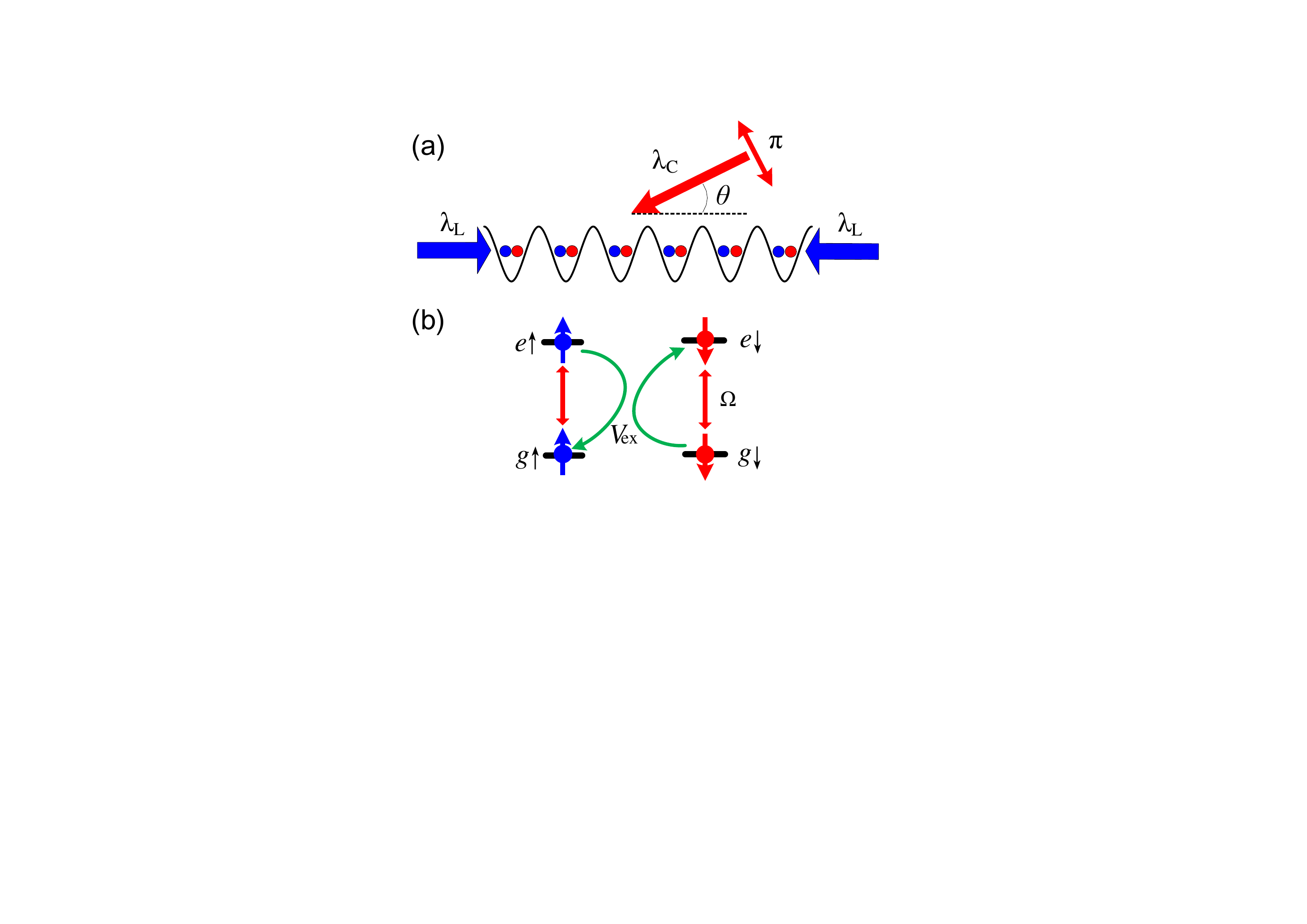}
\caption{(a) Schematics of the experimental setup. The ultracold
alkaline-earth-like atoms are trapped in a 1D optical lattice, which is
generated by a pair of counter-propagating lasers with the \textquotedblleft
magic\textquotedblright\ wavelength $\protect\lambda _{\mathrm{L}}$, such
that states in the clock-state manifolds $^1S_0$ and $^3P_0$ are subject to
the same lattice potential. An ultranarrow $\protect\pi $-polarized clock
laser with a wavelength $\protect\lambda _{\mathrm{C}}$ drives a
single-photon transition between the clock-state manifolds. By introducing
an angle $\protect\theta $ between the clock laser and that generating the
optical lattice, the photon recoil momentum becomes $k_{\mathrm{C}}=2\protect%
\pi \cos \protect\theta /\protect\lambda _{\mathrm{C}}$ and can be tuned
experimentally. (b) Energy levels considering two nuclear spin states $%
\left\vert \uparrow \right\rangle $ and $\left\vert \downarrow \right\rangle
$ in each manifold. While states in the $^1S_0$ and $^3P_0$ manifolds are
coherently coupled by the spin-conserving clock laser (the red curves), the
interaction in the clock-state manifolds can couple distinct spin states in
different orbitals (the green curves).}
\label{experimental}
\end{figure}

We consider a similar setup as in the recent experiment on synthetic SOC in
optical lattice clocks~\cite{clock173Yb}. As shown in Fig.~\ref{experimental}%
(a), a pair of counter-propagating laser with the \textquotedblleft
magic\textquotedblright\ wavelength $\lambda _{\mathrm{L}}=2\pi /k_{\mathrm{L%
}}$ is used to generate a one-dimensional (1D) optical lattice potential $V_{%
\mathrm{L}}\left( x\right) =V_{x}\cos ^{2}(k_{\mathrm{L}}x)$, where $V_{x}$
is the lattice depth. The synthetic SOC is implemented by an ultra-narrow $%
\pi $-polarized clock laser with a wavelength $\lambda _{\mathrm{C}}$, which
drives a single-photon transition between the clock states with the same
nuclear spins. Due to the existence of the angle $\theta $ between the wave
vector of the clock laser and the alignment of the 1D optical lattice [see
Fig.~\ref{experimental}(a)], the momentum transfer becomes $k_{\mathrm{C}%
}=2\pi \cos \theta /\lambda _{\mathrm{C}}$ \cite{clock173Yb}. A key
ingredient in this system is the inter-orbital spin-exchange interaction
\cite{ofr1,ofr2,ofr3}, as shown by the green curves in Fig.~\ref%
{experimental}(b). The full Hamiltonian is written as ($\hbar=1$ hereafter)
\begin{equation}
\hat{H}_{\mathrm{T}}=\hat{H}_{\mathrm{L}}+\hat{H}_{\mathrm{C}}+\hat{H}_{%
\mathrm{I}}  \label{HT}
\end{equation}%
with
\begin{equation}
\hat{H}_{\mathrm{L}}=\sum_{\alpha \sigma }\int dx\hat{\psi}_{\alpha \sigma
}^{\dag }\left( x\right) [-\frac{\nabla ^{2}}{2m}+V_{\mathrm{L}}\left(
x\right) ]\hat{\psi}_{\alpha \sigma }\left( x\right) ,  \label{HL}
\end{equation}%
\begin{equation}
\hat{H}_{\mathrm{C}}=\frac{\Omega _{\text{R}}}{2}\sum_{\sigma }\int dx[\hat{%
\psi}_{g\sigma }^{\dag }\left( x\right) e^{ik_{\mathrm{C}}x}\hat{\psi}%
_{e\sigma }\left( x\right) +\mathrm{H.c.}],  \label{HC}
\end{equation}%
\begin{eqnarray}
\hat{H}_{\mathrm{I}} &\!\!\!\!\!=\!\!\!\!\!&\frac{g_{\pm }}{2}\!\!\int
dx[\Psi _{g\uparrow }^{\dag }\Psi _{e\downarrow }^{\dag }\mp \Psi
_{g\downarrow }^{\dag }\Psi _{e\uparrow }^{\dag }]\left[ \Psi _{e\downarrow
}\Psi _{g\uparrow }\mp \Psi _{e\uparrow }\Psi _{g\downarrow }\right]  \notag
\\
&+&g_{-}\int dx\left[ \Psi _{g\uparrow }^{\dag }\Psi _{e\uparrow }^{\dag
}\Psi _{e\uparrow }\Psi _{g\uparrow }\!+\!\Psi _{g\downarrow }^{\dag }\Psi
_{e\downarrow }^{\dag }\Psi _{e\downarrow }\Psi _{g\downarrow }\right] ,
\label{Hi}
\end{eqnarray}%
where $\alpha =\left\{ g,e\right\} $ is the orbit index, $\sigma =\left\{
\uparrow ,\downarrow \right\} $ is the spin index, $\Psi _{\alpha \sigma }$
and $\Psi _{\alpha \sigma }^{\dag }$ are the corresponding field operators, $%
\Omega _{\text{R}}$ is the Rabi frequency of the clock laser, $g_{\pm }$ are
the 1D interaction strengths~\cite{ren1,ren2}, and $\mathrm{H.c.}$ is the
Hermitian conjugate. Note that in writing down Hamiltonian (\ref{HT}),
we only consider four nuclear spin states from the $|g\rangle$ and $|e\rangle
$ manifolds. In principle, the other nuclear spin states can be shifted away
by imposing spin-dependent laser shifts~\cite{AEnew8}.

When the 1D optical lattice is deep enough and $\Omega _{\text{R}}$ is not
too large \cite{highband1,highband2,highband4}, we may take the single-band
approximation and write down the corresponding tight-binding model
\begin{eqnarray}
\hat{H}_{\mathrm{TB}} &\!\!=\!\!&-t\!\!\sum_{<i,j>,\alpha \sigma }\hat{c}%
_{i\alpha \sigma }^{\dag }\hat{c}_{j\alpha \sigma }\!+\!\frac{\Omega }{2}%
\sum_{j,\sigma }(e^{i\phi j}\hat{c}_{jg\sigma }^{\dag }\hat{c}_{je\sigma
}\!+\!\mathrm{H.c.})  \notag \\
&&+U\sum_{j}(\hat{n}_{jg\uparrow }\hat{n}_{je\downarrow }+\hat{n}%
_{jg\downarrow }\hat{n}_{je\uparrow })+U_{0}\sum_{j\sigma }\hat{n}_{jg\sigma
}\hat{n}_{je\sigma }  \notag \\
&&+V_{\mathrm{ex}}\sum_{j}(\hat{c}_{jg\uparrow }^{\dag }\hat{c}%
_{je\downarrow }^{\dag }\hat{c}_{je\uparrow }\hat{c}_{jg\downarrow }+\mathrm{%
H.c.}),  \label{HTB}
\end{eqnarray}%
where $\hat{c}_{j\alpha \sigma }$ ($\hat{c}_{j\alpha \sigma }^{\dag }$) is
the annihilation (creation) operator for atoms on the $i$th site of the
$\alpha $ orbital and the spin $\sigma $, $\hat{n}_{j\alpha \sigma }=\hat{c}%
_{j\alpha \sigma }^{\dag }\hat{c}_{j\alpha \sigma }$. The spin-conserving hopping rate $t${$=\left\vert \int
dx{w}^{(j)}\left[ -\frac{\nabla ^{2}}{2m}+V_{\mathrm{L}}\left( x\right) %
\right] w^{(j+1)}\right\vert $, with $w^{(j)}$ being the lowest-band Wannier
function on the $j$th site of the lattice potential $V_{\mathrm{L}}\left(
x\right) $,}. The spin-flipping hopping rate $\Omega $ {$=\Omega _{%
\text{R}}\int dxw^{(j)}e^{ik_{\mathrm{C}}x}w^{(j)}$ }. $\phi =\frac{1}{2}k_{\mathrm{C}}\lambda _{\mathrm{L}}=\pi
\lambda _{\mathrm{L}}\cos \theta /\lambda _{\mathrm{C}}$ is the synthetic
magnetic flux per plaquette induced by the SOC, $U=\frac{1}{2}\left(
g_{+}+g_{-}\right) \int dxw^{(j)}w^{(j)}w^{(j)}w^{(j)}$ and $U_{0}=${$%
g_{-}\int dxw^{(j)}w^{(j)}w^{(j)}w^{(j)}$} are the inter-orbital
density-density interaction strengths with the same and different nuclear
spins, respectively. $V_{\mathrm{ex}}=\frac{1}{2}\left(
g_{-}-g_{+}\right) \int dxw^{(j)}w^{(j)}w^{(j)}w^{(j)}$ is the inter-orbital
spin-exchange interaction strength. Hamiltonian (\ref{HTB}) has the
advantage that all parameters can be tuned independently. For example, $t$
can be controlled by the depth of the optical lattice potential, $\Omega $
and $\phi $ can be controlled by the Rabi frequency and the angle of the
clock laser, respectively, and $\{V_{\mathrm{ex}}$, $U$, $U_{0}\}$ can be
tuned through the orbital Feshbach resonance~\cite{ren1,ofrexp1,ofrexp2} or
the confinement induced resonance~\cite{CIR,ren2}. In the following, we take
$U_0=V_{\mathrm{ex}}+U$, which is dictated by the scattering parameters of $%
^{173}$Yb atoms~\cite{CIR,ren2}.

\begin{figure}[tb]
\centering
\includegraphics[width =8.5cm]{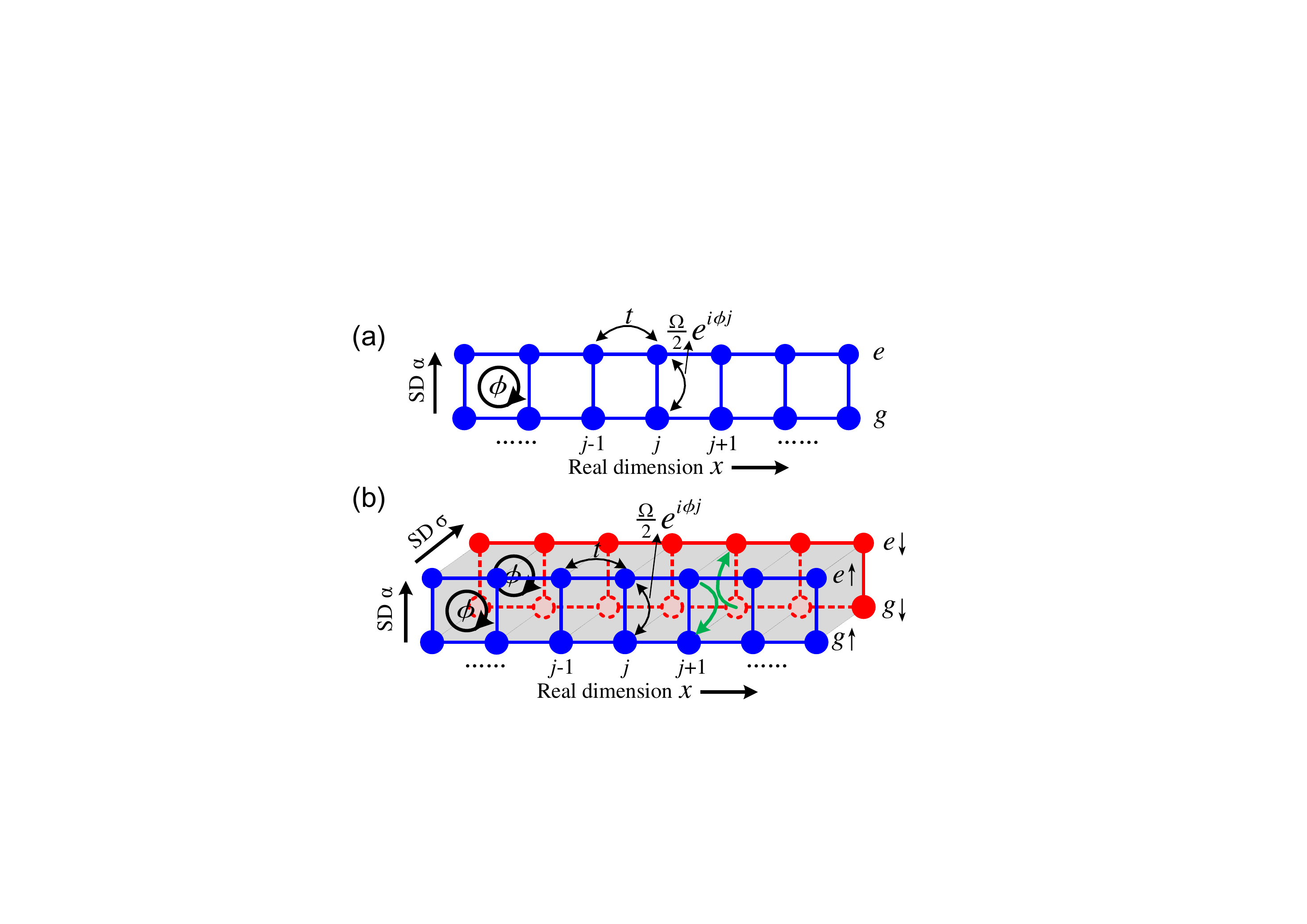}
\caption{(a) A two-leg synthetic ladder with a synthetic magnetic flux $%
\protect\phi =\protect\pi \protect\lambda _{\mathrm{L}}/\protect\lambda _{%
\mathrm{C}}\cos \protect\theta $ in each plaquette. The orbital states can
be treated as an effective synthetic dimension denoted as SD $\protect\alpha
$. The two ladders along the synthetic dimension SD $\protect\sigma $ are
identical and decoupled. (b) A pair of two-leg synthetic ladders (i.e., $%
\left\vert g\uparrow ,e\uparrow \right\rangle $ and $\left\vert g\downarrow
,e\downarrow \right\rangle $), with the same flux, are coupled by the
inter-orbital spin-exchange interaction (green curves). In this lattice,
there are two-direction synthetic dimensions, SD $\protect\alpha $ and SD $%
\protect\sigma $.}
\label{lattice}
\end{figure}

From the tight-binding Hamiltonian (\ref{HTB}), it is clear that if we drop
the interaction terms and map the electronic ($\alpha$) and the spin ($%
\sigma $) states onto effective lattice sites along two different synthetic
dimensions, the non-interacting tight-binding model describes a pair of
two-leg ladders. We label the two synthetic dimensions as the orbit $\alpha $%
- and the spin $\sigma $-directions, respectively, while the optical lattice
lies along the $x$-direction. We may then denote the synthetic dimensions as SD $%
\alpha $ and SD $\sigma $, respectively. As illustrated in Fig.~\ref{lattice}%
, the pair of ladders each lie within the $(x,\alpha)$ plane with the legs
of both ladders along the $x$-direction. The rung tunneling in each ladder
is facilitated by the SOC, which also induces uniform magnetic flux in each
plaquette of the ladder. In the absence of interactions, the ladders are not
coupled, as different spin states are independent on the single-body level.
However, the inter-orbital spin-exchange interaction effectively couples the
ladders together [see Fig.~\ref{lattice}(b)], which, as we will show later,
induce inter-ladder currents along the $\sigma$-direction.

An important property here is the current along the legs and the rungs of
the ladder. Local and average currents along the $x$-direction can be define
as~\cite{Boseladder1,Boseladder2,helicalliquidnc},
\begin{align}
J_{j,\alpha \sigma }^{\parallel }& =i\left( \hat{c}_{j+1\alpha \sigma
}^{\dag }\hat{c}_{j\alpha \sigma }-\hat{c}_{j\alpha \sigma }^{\dag }\hat{c}%
_{j+1\alpha \sigma }\right) , \\
J_{\alpha \sigma }^{\parallel }& =\frac{1}{L}\sum_{j}J_{j,\alpha \sigma
}^{\parallel }.
\end{align}%
Similarly, currents along the $\alpha $-direction can be defined as~\cite%
{Boseladder2},
\begin{align}
J_{j,\alpha }^{\perp }& =i(e^{i\phi j}\hat{c}_{je\sigma }^{\dag }\hat{c}%
_{jg\sigma }-e^{-i\phi j}\hat{c}_{jg\sigma }^{\dag }\hat{c}_{je\sigma }), \\
J_{\alpha }^{\perp }& =\frac{1}{L}\sum_{j}\left\vert J_{j,\alpha }^{\perp
}\right\vert ,
\end{align}%
Finally, we also define currents along the $\sigma $-direction as
\begin{align}
J_{j,\sigma }^{\perp }& =i(\hat{c}_{j\alpha \downarrow }^{\dag }\hat{c}%
_{j\alpha \uparrow }-\hat{c}_{j\alpha \uparrow }^{\dag }\hat{c}_{j\alpha
\downarrow }), \\
J_{\sigma }^{\perp }& =\frac{1}{L}\sum_{j}\left\vert J_{j,\sigma }^{\perp
}\right\vert .
\end{align}

In the following discussions, we adopt the DMRG formalism to calculate the
ground state of the system, from which we characterize currents and density
correlation functions. For the numerical calculation, we have considered
length of chain $L$ up to 32 sites. We keep the maxstates $m = 200$ and
achieve truncation errors of $10^{-10}$. We mainly consider the case of half
filling, i.e., $n=N/(2L)=1$, where $L$ is the length of chain and $N$ is the
total number of atoms.\newline

\section{Phases and currents in the non-interacting case}

\label{Phases and currents in the non-interacting case}

\begin{figure}[t]
\centering
\includegraphics[width = 3.5in]{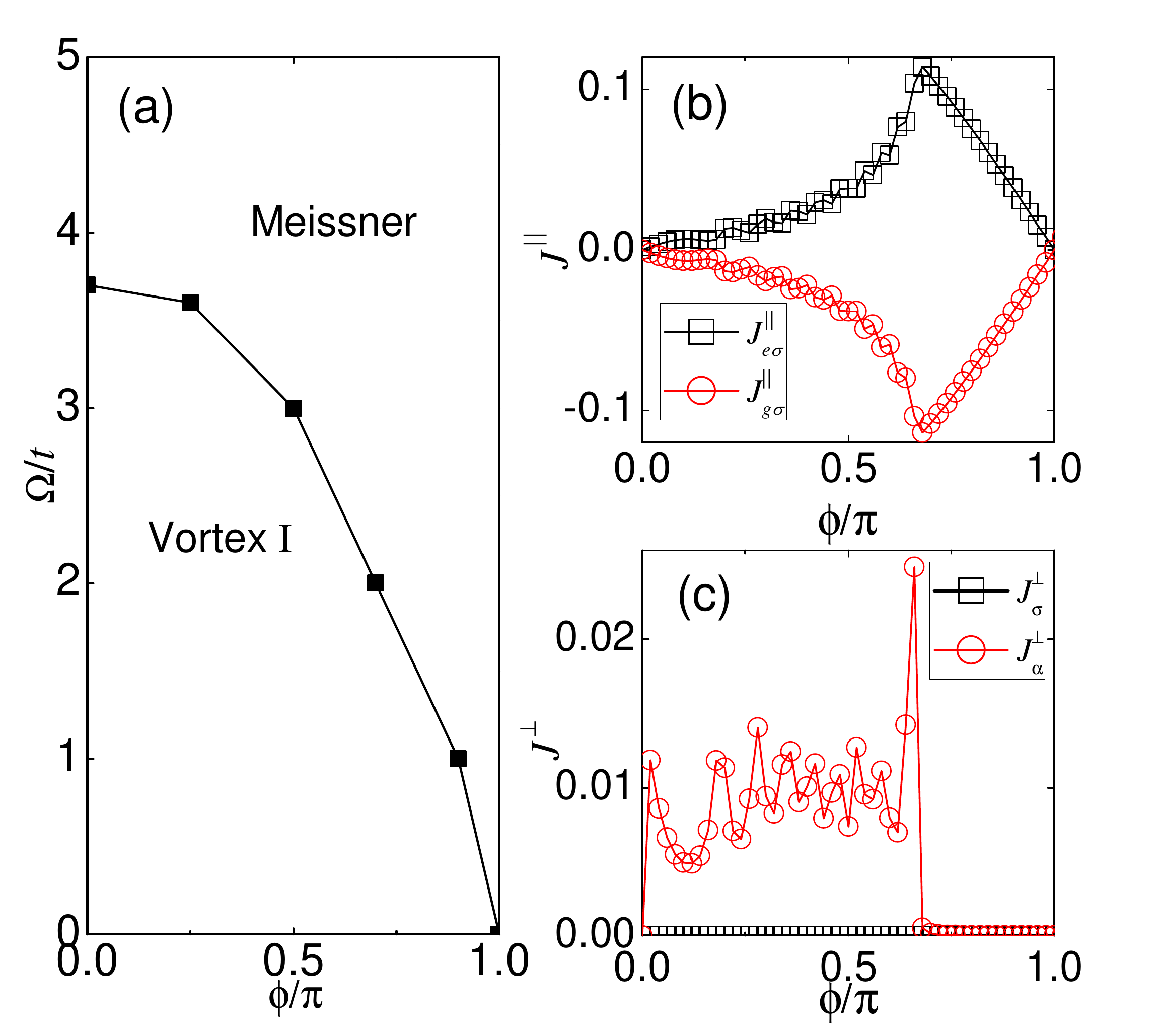} \hskip -0.0cm
\caption{(a) Phase diagram in the $(\Omega,\protect\phi)$ plane. The
currents (b) $J^{\parallel }$ and (c) $J^{\perp }$ as functions of $\protect%
\phi /\protect\pi $. In all subfigures, $U=U_{0}=V_{\mathrm{ex}}=0$ and $n=1$%
, and (b) and (c) have the other parameter $\Omega /t=2$.}
\label{Phase diagram_withoutinteraction}
\end{figure}

We first discuss the ground-state phases of the system in the absence of
interactions. In this case, different spin states are decoupled, and we may
identify a pair of two-leg ladders, as illustrated in Fig.~\ref{lattice}.
From our numerical calculations, we find that only the Meissner and the
vortex states appear in the ground-state phase diagram shown in Fig.~\ref%
{Phase diagram_withoutinteraction}(a). Typically, when the synthetic flux is
small, the ground state is the so-called Meissner state, where the edge
currents $J_{g\sigma }^{\parallel }$ and $J_{e\sigma }^{\parallel }$ flow in
opposite directions along the two legs of each ladder~\cite%
{clock173Yb,RamanRb,Boseladder1}, as shown in Fig.~\ref{Phase
diagram_withoutinteraction}(b). Upon increasing the flux above a critical
value, the ground state features a vortex state with rung currents and
vortex lattice structures in the bulk of each ladder, i.e., in the $%
(x,\alpha )$ plane. The existence of this so-called Vortex I state is
confirmed in Fig.~\ref{Phase diagram_withoutinteraction}(c), where nonzero $%
J_{\alpha }^{\perp }$ in the Vortex I state regime indicates a finite
current along the rungs in the $\alpha $-direction for each ladder. The
current $J_{\sigma }^{\perp }$ remains zero in the non-interacting case,
which is consistent with the picture of two independent ladders of different
spins. We also note that in the non-interacting case, there are no
density-ordered phases.\newline

\section{Impact of interactions on the Meissner state}

\label{Impact of interactions on the Meissner state}

\begin{figure}[t]
\centering
\includegraphics[width = 3.5in]{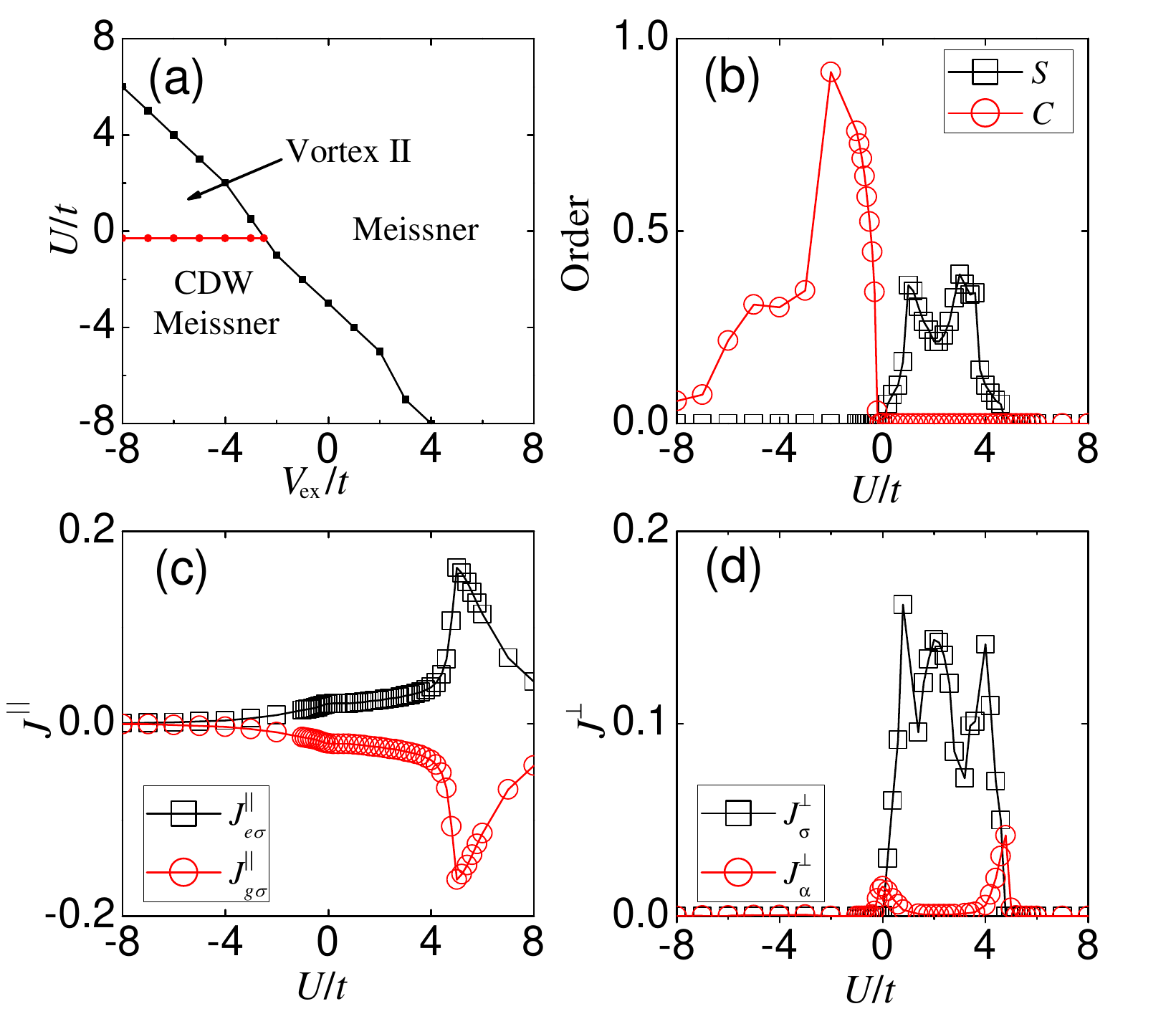} \hskip -0.0cm
\caption{(a) Phase diagram in the $(U,V_{\mathrm{ex}})$ plane. (b) The SDW
order $S$ and the CDW order $C$ as well as the currents (c) $J^{\parallel }$
and (d) $J^{\perp }$ as functions of $U/t$. In all subfigures, $\Omega /t=4$%
, $\protect\phi /\protect\pi =0.75$, and $n=1$, and (b)-(d) have the other
parameter $V_{\mathrm{ex}}/t=-6$.}
\label{Phase diagram_interaction_a}
\end{figure}

We now study the impact of interactions on the Meissner state for $\Omega/t
=4$ and $\phi /\pi =0.75$. As the interactions are turned on, the system can
undergo phase transitions into exotic vortex states or phases with density
orders. We map out the phase diagram in the $(U,V_{\mathrm{ex}})$ plane,
while fixing other parameters. As shown in Fig.~\ref{Phase
diagram_interaction_a}(a), the phase diagram consists of three different
phases: a simple Meissner state, a Meissner state with CDW, and an exotic
vortex state with SDW, which we label as Vortex II state. The simple
Meissner state resembles the Meissner state in the non-interacting case with
chiral edge currents and no density-orders in the bulk. The CDW Meissner
state features chiral edge currents as well as finite CDW density
correlations in the bulk. The most interesting state here is the Vortex II
state, which features finite SDW correlations as well as currents and vortex
lattice structures in the $(x,\sigma)$ plane.

\begin{figure}[t]
\centering
\includegraphics[width=3.5in]{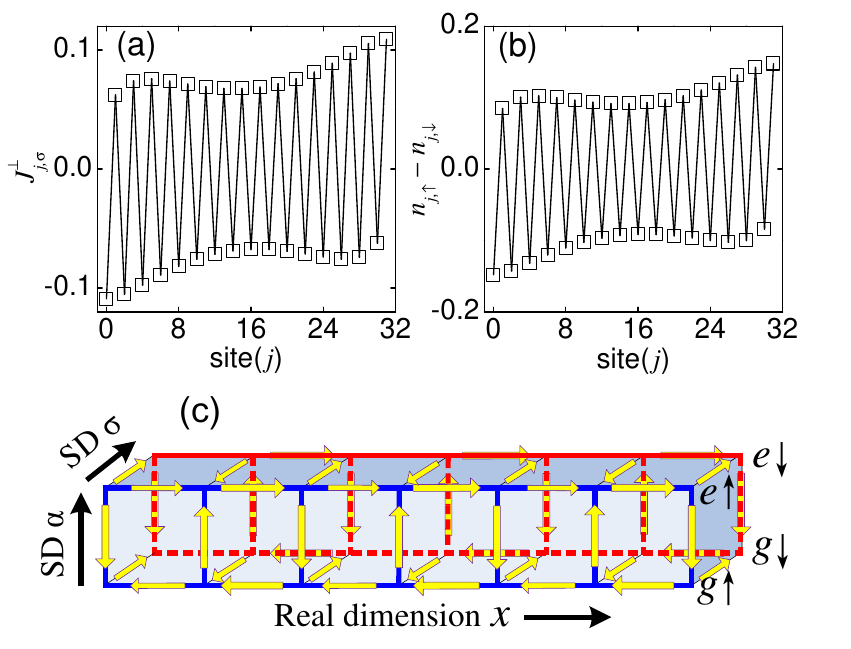} \hskip -0.0cm
\caption{(a) The currents $J^{\perp }$ and (b) density profiles $%
n_{j,\uparrow }-n_{j,\downarrow }$ for different sites. (c) Sketch of
currents of the Vortex II state. In (a) and (b), $\Omega /t=4$, $\protect%
\phi /\protect\pi =0.75$, $V_{\mathrm{ex}}/t=-6$, $U/t=2$, and $n=1$.}
\label{Vortex2phase}
\end{figure}

The phase boundaries between these phases can be determined from the CDW and
the SDW correlations, as well as from the currents' calculations. As
illustrated in Fig.~\ref{Phase diagram_interaction_a}(b), for a fixed $V_{%
\mathrm{ex}}/t=-6$, the CDW order
\begin{equation}
{C=\frac{1}{2L}\sum_{j}(-1)^{j}n_{j}\ }  \label{CDW}
\end{equation}%
has a finite value in the range $-8<U/t<0$, while the SDW order
\begin{equation}
S=\frac{1}{{L}}\sum_{j}(-1)^{j}(n_{j,\uparrow }-n_{j,\downarrow })
\label{SDW}
\end{equation}%
has a finite value for $0<U/t<4.7$. On the other hand, while the edge
currents $J_{\alpha \sigma }^{\parallel }$ are always finite and opposite in
directions for different sites in the $\alpha $ direction, the currents $%
J_{\sigma }^{\perp }$ and $J_{\alpha }^{\perp }$ only exist within a range,
as shown in Figs.~\ref{Phase diagram_interaction_a}(c) and \ref{Phase
diagram_interaction_a}(d). In particular, the non-vanishing $J_{\sigma
}^{\perp }$ indicates inter-ladder currents and vortices in the $(x,\sigma )$
plane. To further characterize the Vortex II state, in Figs.~\ref%
{Vortex2phase}(a) and \ref{Vortex2phase}(b), we show the spatial
distribution of the inter-ladder currents as well as the spin density.
Apparently, the SDW and the vortex lattice structure in the $(x,\sigma )$
plane is due to the interplay of the inter-orbital spin-exchange interaction
and the synthetic magnetic flux in the $(x,\alpha )$ plane, as shown in Fig.~%
\ref{Vortex2phase}(c).\newline

\section{Impact of interactions on the vortex state}

\label{Impact of interactions on the vortex state}

\begin{figure}[t]
\centering
\includegraphics[width = 3.5in]{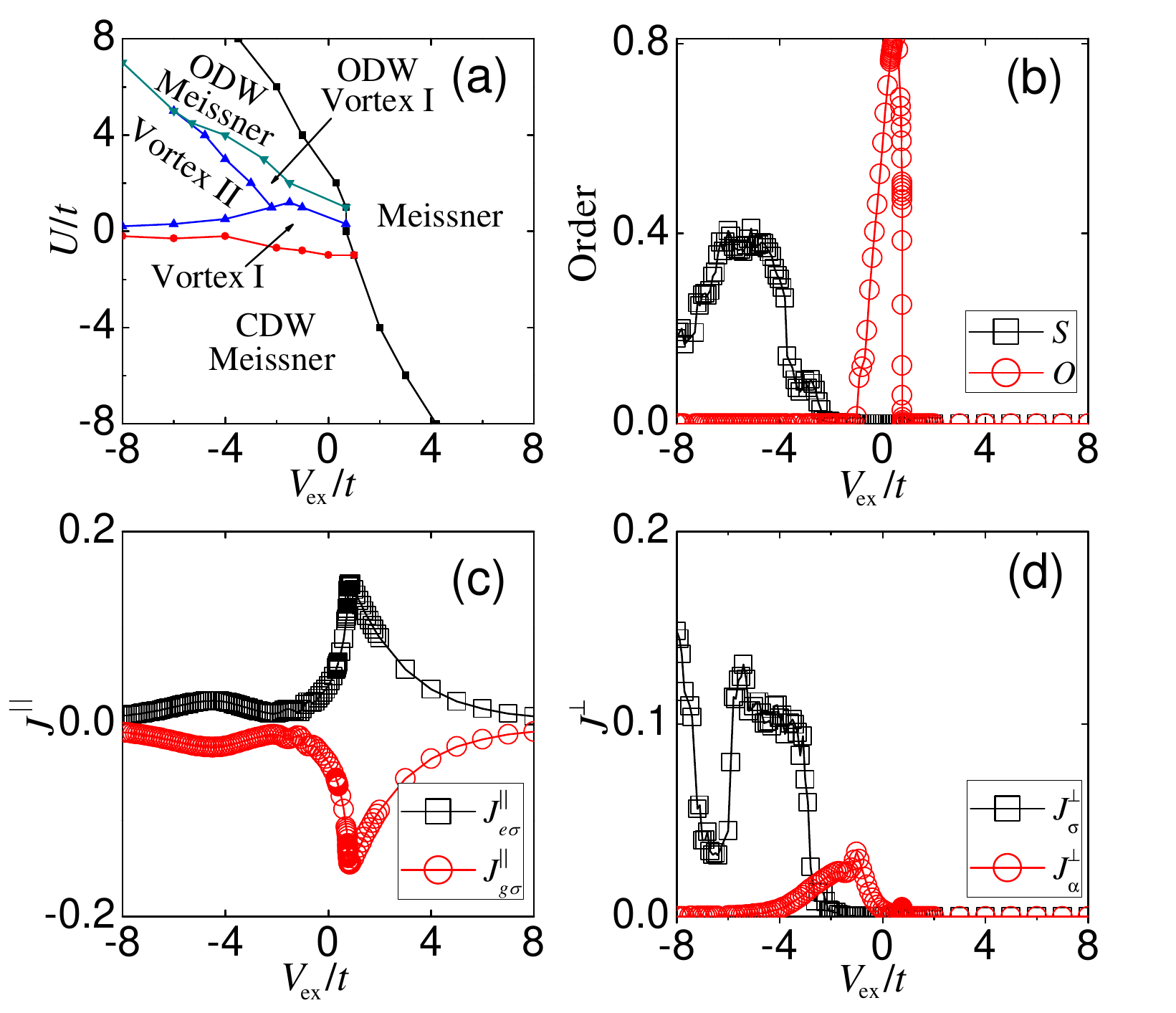} \hskip -0.0cm
\caption{(a) Phase diagram in the $(U,V_{\mathrm{ex}})$ plane. (b) The SDW
order $S$ and the ODW order $O$ as well as the currents (c) $J^{\parallel }$
and (d) $J^{\perp }$ as functions of $V_{\mathrm{ex}}/t$. In all subfigures,
$\Omega /t=2$, $\protect\phi /\protect\pi =0.25$, and $n=1$, and (b)-(d)
have the other parameter $U/t=1 $.}
\label{Phase diagram_interaction_b}
\end{figure}

In this section, we study the impact of interactions on the vortex states.
In Fig.~\ref{Phase diagram_interaction_b}, we map out the phase diagram in
the $(U,V_{\mathrm{ex}})$ plane for $\Omega /t=2$ and $\phi /\pi =0.25$. In
the absence of interactions, the system is in the vortex state. With
interactions, the system can undergo phase transitions into various
different phases. As shown in Fig.~\ref{Phase diagram_interaction_b}(a),
besides the Vortex II state and the CDW Meissner state, several other exotic
phases emerge in the phase diagram. While the Vortex I resembles the vortex
state in the absence of interactions, interesting phases with density
correlations in the orbital channel appear, which can be further
differentiated by their currents flows as the ODW Meissner state and the ODW
Vortex I state, where the vortex occurs in the $(x,\alpha )$ plane, together
with density-wave orders in the orbital channel. Here the ODW order is
defined as
\begin{equation}
{O=\frac{1}{{L}}\sum_{j}(-1)^{j}(n_{j,g}-n_{j,e}).}  \label{ODW}
\end{equation}

In Figs.~\ref{Phase diagram_interaction_b}(b)-\ref{Phase
diagram_interaction_b}(d), we plot the currents and the density wave orders
as functions of $V_{\mathrm{ex}}/t$ for $U/t=1$. For $V_{\mathrm{ex}}/t\in
\lbrack -8,-2.06]$, the SDW order $S$ has a finite value, the ODW order
vanishes with $O=0$, the currents $J_{\alpha \sigma }^{\parallel }>0$ and $%
J_{\sigma }^{\perp }>0$, the corresponding phase is the Vortex II state.
When $V_{\mathrm{ex}}/t\in \lbrack -2.06,-1]$, the SDW order $S=0$, the ODW
order $O=0$, the currents $J_{\alpha \sigma }^{\parallel }>0$, $J_{\alpha
}^{\perp }>0$, and $J_{\sigma }^{\perp }=0$, the corresponding phase is the
Vortex I state. When $V_{\mathrm{ex}}/t\in $ $[-1,0.77]$, the SDW order $S=0$%
, the ODW order $O>0$, the currents $J_{\alpha \sigma }^{\parallel }>0$, $%
J_{\alpha }^{\perp }>0$, and $J_{\sigma }^{\perp }=0$, the corresponding
phase is the ODW Vortex I state. When $V_{\mathrm{ex}}/t\in $ $[0.77,8]$,
the SDW order $S=0$, the ODW order $O=0$, the currents $J_{\alpha \sigma
}^{\parallel }>0$, $J_{\alpha }^{\perp }=0$, and $J_{\sigma }^{\perp }=0$,
the corresponding phase is the Meissner state.

According to Hamiltonian (\ref{HTB}), there are three different
interaction parameters $U$, $V_{\mathrm{ex}}$ and $U_{0}$. It is
straightforward to see that the CDW order is favored for the attractive
interactions ($U<0$, $U_{0}<0$, $V_{\mathrm{ex}}<0$); and that the SDW order
is favored for the attractive inter-orbital spin-preserving interaction ($%
U_{0}<0$). This is because the CDW order can decrease the interaction energy of all the attractive on-site interactions. While the SDW order mainly decreases the energy of the attractive inter-orbital spin-preserving interactions. These are consistent with previous studies on the SU(2) ladder
systems \cite{phaseEPL,twoorbitSUn1}. Note that for the phase diagrams in
Figs.~\ref{Phase diagram_interaction_a} and \ref{Phase diagram_interaction_b}, we have fixed $U_{0}=V_{\mathrm{ex}}+U$. In the phase
diagrams, the CDW state becomes unstable against the Vortex II state when $U$
is not sufficiently negative. Apparently, the competition between the CDW
and the SDW orders is driven by the interactions associated with $U$ and $%
U_{0}$, which is eventually determined by the relative values of $U$ and $V_{%
\mathrm{ex}}$. On the other hand, the emergence of the ODW order in Fig.~\ref{Phase diagram_interaction_b}
can be understood as a configuration in which the repulsive energies are
minimized in the case of $U>0$ and $U_{0}>0$. A subtlety here is the impact
of the magnetic flux in the $x$-$\alpha $ plane on the density ordered
phases. From numerical analysis, we see that an increase of magnetic flux
can lead to a stronger competition between different density-ordered phases,
which gives rise to a richer phase diagram. In any case, we emphasize that
the exotic Vortex II state with the SDW order is always robust in the
negative $U_{0}$ limit.\newline

\section{Detection}

\label{Discussions}

Given the rich phase diagram discussed above, a natural question is how to
detect them experimentally. In general, the different phases of the system
are characterized by their chiral edge currents as well as the density
correlations in different channels such as CDW, ODW and SDW. Here, the
density orders can be probed by state-selective measurements of density
distributions. While the CDW order can be identified by oscillations of the
total density distribution from site to site, the ODW and the SDW orders can
be identified, respectively, by oscillations of the density distribution of
a given orbital ($|g\rangle $ or $|e\rangle $) or of a given spin ($%
|\!\!\uparrow \rangle $ or $|\!\!\downarrow \rangle $).
For the detection of the Meissner and the vortex states, one can in
principle follow the approach in Ref.~\cite{Boseladder2}, where, by
projecting the wave function into isolated double wells along each leg, the
chiral currents can be calculated from the oscillatory density dynamics in
the double wells. The vortex state can be identified either from the
variation of the chiral currents, as the maximum of the chiral currents
appear at the phase boundary between the Meissner and the vortex states [see
Figs.~\ref{Phase diagram_interaction_a}(a) and \ref{Phase
diagram_interaction_b}(a)]. Alternatively, one should also identify the
Vortex I and the Vortex II states, respectively, from the relative phase
between the oscillations of different orbital and spin states. The relative
phase should be $\pi $ in the case of the Meissner state, and smaller than $%
\pi $ in the case of the vortex states~\cite{Boseladder2}.

\section{Conclusion}

\label{Conclusion}

We show that by implementing synthetic SOC in alkaline-earth-like atoms, one
naturally realizes multiple two-leg ladders with uniform synthetic flux. As
interactions couple different ladders together, the system features a rich
phase diagram. In particular, we demonstrate the existence of an
interaction-induced vortex state, which possesses SDW in the spin channel.
The different phases can be experimentally detected based on their
respective properties. As many phases in the phase diagram
simultaneously feature density order and edge or bulk currents, a potential
experimental challenge lies in the efficient detection of the various
phases. This is particularly so for the exotic interaction-induced Vortex II
state, whose SDW order as well as the bulk currents along SD $\sigma$
require spin-selective detections. Nevertheless, with the state of the art
quantum control over the clock states in alkaline-earth-like atoms, we expect that these
challenges can be overcome with existing experimental techniques. Our
results reveal the impact of interactions on spin-orbit coupled systems, and
are particularly relevant to the on-going exploration of spin-orbit coupled
optical lattice clocks.\newline

\section*{Acknowledgments}

This work is supported partly by the National Key R\&D Program of China
under Grants No.~2017YFA0304203 and No.~2016YFA0301700; the NKBRP under
Grant No.~2013CB922000; the NSFC under Grants No.~60921091, No.~11374283,
No.~11434007, No.~11422433, No.~11522545, and No.~11674200; ``Strategic
Priority Research Program(B)'' of the Chinese Academy of Sciences under
Grant No.~XDB01030200; the PCSIRT under Grant No.~IRT13076; the FANEDD under
Grant No.~201316; SFSSSP; OYTPSP; and SSCC. J.-S. P. acknowledges support
from National Postdoctoral Program for Innovative Talents of China under
Grant No.~BX201700156.

\end{document}